\documentclass{raa}
\usepackage{graphicx,times}
\usepackage{natbib}
\usepackage{amssymb,amsmath}
\bibpunct{(}{)}{;}{a}{}{,}

\usepackage[a4paper=true,dvipdfm=true,pagebackref=true]{hyperref}
\hypersetup{pdftitle = The title of my PDF, pdfauthor = My name, pdfsubject= The subject, pdfkeywords = keyword1 keyword2 keyword3}
\hypersetup{colorlinks = true, linkcolor = green, anchorcolor = red, citecolor = blue, filecolor = red, pagecolor = red, urlcolor = red}
\def\fm{\hbox{$.\!\!^m$}}
\def\fs{\hbox{$.\!\!^s$}}
\def\degr{\hbox{$^\circ$}}

\begin{document}%

\title{Physical and Geometrical Parameters of CVBS X: The Spectroscopic Binary  Gliese 762.1}

\volnopage{ {\bf 2016} Vol.\ {\bf X} No. {\bf XX}, 000--000}
   \setcounter{page}{1}

   \author{Suhail G. Masda\inst{1}, Mashhoor A. Al-Wardat\inst{2}, Ralph Neuh\"{a}user\inst{1}, Hamid M. Al-Naimiy\inst{3}  }

\institute{ Astrophysikalisches Institut und Universit\"{a}ts-Sternwarte Jena, FSU Jena, 07745 Jena, Germany
        \and  Physics Department, Faculty of Science,  Al al-Bayt University, PO Box: 130040, Mafraq, 25113 Jordan; {\it mwardat@aabu.edu.jo}   \\
        \and  Applied Physics and Space Science Department, University of Sharjah PO Box: 27272 Sharjah, United Arab Emirates \\
\vs \no
   {\small Received 2015 December 3; accepted 2016 March 10}
}

\abstract{
We present the physical and geometrical parameters of the individual components of the close visual double-lined spectroscopic binary system
Gliese 762.1, which were estimated using Al-Wardat's complex method for analyzing close visual binary systems.  The estimated parameters of the individual components of the system are as follows: radius $R_{A}=0.845\pm0.09 R_\odot$, $R_{B}=0.795\pm0.10 R_\odot$, effective temperature $T_{\rm eff}^{A}
=5300\pm50$\,K, $T_{\rm eff}^{B} =5150\pm50$\,K, surface gravity log $g_{A}=4.52\pm0.10$,
log $g_{B}=4.54\pm0.15$ and luminosity  $L_A=0.51\pm0.08 L_\odot$, $L_B=0.40\pm0.07L_\odot$. New orbital elements are presented with a semi-major axis of $0.0865 \pm 0.010 $ arcsec using the Hippracos parallax $\pi=58.96\pm0.65$ mas, and  an accurate total mass and individual masses of the system are  determined as $M=1.72\pm0.60M_\odot$, $M_A=0.89 \pm0.08M_\odot$ and $M_B=0.83 \pm0.07M_\odot$. Finally, the spectral types and luminosity classes of both components are assigned as K0V and K1.5V for the primary and secondary components respectively, and their positions on the H-R diagram and evolutionary tracks are given.
}

\authorrunning{S. G. Masda et al. }            
   \titlerunning{Parameters of CVBS XI: Gliese 762.1}  
\maketitle%

\section{INTRODUCTION }
The importance of the study of binary stars arises from the fact that more than 50\% among nearby solar-type main-sequence stars  are binary or multiple stellar systems (the fraction is 42\% among nearby M stars~\cite{1991A&AS...88..281D}) and several astronomical phenomena occur only in binary stars.
 They provide a source of direct measurements of stellar parameters or
galactic quantities.  Stellar physics needs masses, luminosities and radii
obtained through the studies of binary stars. Galactic physics benefits also from these studies, e.g.
the galactic potential can be tested using wide binaries, and the chemical evolution depends
on binaries through the supernovae Ia process ~\citep{2002EAS.....2..155A}. The mass-luminosity relation of low-mass main-sequence  stars in the solar neighborhood are known with much lower accuracy than those of massive early-type stars \cite{2002A&A...385...87B, 2007A&A...464..635B, 1999A&A...351..619F}, this requires precise determination of their masses and luminosities.

The close visual binary stars (CVBS) are visually close enough  to be resolved except by special techniques like speckle interferometry or by deducing their duplicity using high resolution spectroscopy. So, the case is a bit complicated and needs indirect methods to estimate their physical and geometrical parameters.

Combining  observational measurements with stellar theoretical models is the most powerful indirect method to analyze such binary and multiple systems. This was implemented in Al-Wardat's  complex method ~\citep{2007AN....328...63A},  which combines magnitude difference measurements of speckle interferometry,  entire spectral energy distribution (SED) of spectrophotometry and radial velocity measurements, all along with atmospheres modeling to estimate the individual physical parameters. In coordination with these physical parameters, the geometrical parameters and their errors are calculated using a modern version of  Tokovinin's ORBITX   program, which depends on the standard least-squares method  \cite{1992ASPC...32..573T}.

The method was firstly introduced by~\cite{2002BSAO...53...51A, 2007AN....328...63A}(henceforth paper I and paper II in this series respectively), where it was applied to the analysis of the quadruple hierarchical system ADS11061 and the two binary systems COU1289 and COU1291 .
Later on, the method was successfully applied to several solar-type and subgiant binary systems: Hip11352, Hip11253 and Hip689 ~\cite{2009AN....330..385A, 2009AstBu..64..365A, 2012PASA...29..523A} (henceforth papers III, IV and V in this series respectively).

 The method was then developed to a complex one by combining the physical solution with the geometrical one represented by the orbital solution of the system and applied to the systems: HD25811, HD375, Gliese 150.5 and HD6009 ~\cite{2014AstBu..69...58A, 2014AstBu..69..198A, 2014PASA...31....5A, 2014AstBu..69..454A}  (henceforth papers VI, VII, VIII and IX in this series respectively).

In order to be analyzed using Al-Wardat's method, the binary system  should have a magnitude difference measurement, an observational entire SED covering the optical range, and a precise entire optical (UBV) photometrical magnitude measurement. The procedure starts with calculating the individual flux of each component using the magnitude difference with the entire photometrical magnitude, then estimating their preliminary effective temperatures and gravity accelerations in order to build their SED using grids of~\cite{1994KurCD..19.....K} blanketed models (ATLAS9). These two models in their turn are used to build a synthetic entire SED for the system, which compares with the observational one in an iterative way until the best fit is achieved.  Of course there should be a coincidence  between the masses calculated using the physical solution and those calculated using the orbital one, otherwise a new set of parameters would be tested.

This is the tenth paper in this series, which gives a complete analysis of the  CVBS Gliese 762.1  using Al-Wardat's  complex method.

Gliese 762.1   (MCA 56 AB = WDS J19311+5835 = GJ 762.1 = HD 184467 = HIP 95995) was first visually resolved by \cite{1983ApJS...51..309M}. It is a well known double-lined spectroscopic binary (SB2)~\cite{2000A&AS..145..161P}, with an orbital period of 1.35 year \cite{1989PDAO...17....1B}. It shines at an apparent visual magnitude of $ m_v=6\fm60$ and the spectral types of both components are catalogued as K2V and K4V for the primary and secondary components respectively~\cite{2010AJ....139.2308F}.
Hipparcos trigonometric parallax measurement of the system as $\pi=58.96\pm0.65$ mas~\cite{2007A&A...474..653V} places this system at a distance of 16.96 pc, being one of the nearby K-type stars.

Table \ref{tlab2} contains basic data of the system Gliese 762.1 from SIMBAD,  NASA/IPAC, Hipparcos and Tycho Catalogues \cite{1997yCat.1239....0E}.

\begin{table}[!h]
	\begin{center}
		\caption{Basic data of the system Gliese 762.1 from SIMBAD, Hipparcos and Tycho Catalogues} \label{tlab2}
		\begin{tabular}{lcc}\hline\hline
			& GJ 762.1 & source of data
			\\
			\hline
			$\alpha_{2000}$ $\dagger$ & $19^h 31^m 07\fs974$&SIMBAD\\
			$\delta_{2000}$ $\ddagger$&$+58\degr35' 09.''64$&-\\
			HIP& 95995 & -\\
			Sp. Typ. & K1V&-\\
			$E(B-V)^{*}$ & $0.07\pm 0.002$& NASA/IPAC\\
			$A_v^{*}$&$0\fm21$& NASA/IPAC
			\\
			$B_J(Hip)$ & $7\fm46$& Hipparcos
			\\
			$V_J(Hip)$ & $6\fm60$& -
			\\
			$R_J(Hip)$ & $6\fm10$& -\\
			$(B-V)_J(Hip)$ & $0\fm86\pm0.001$& -
			\\
			$(U-B)_{J}$ & $0\fm52\pm0.001$&-
			\\
			$B_T$ & $7\fm71\pm0.006$ & Tycho
			\\
			$V_T$ & $6\fm71\pm0.005$& -
			\\
			$(B-V)_J(Tyc)$ & $0\fm87\pm0.006$& -
			\\
			$\pi_{Hip}$ (mas) & $59.84\pm0.64$ & Hipparcos
			\\
			$\pi_{Tyc}$ (mas) & $58.00\pm2.90$& Tycho
			\\
			$\pi^{**}_{Hip}$  (mas) & $58.96\pm0.65$& New Hipparcos
			\\
			\hline\hline
		\end{tabular}
		\\
		$\dagger$ Right Ascention,
		$\ddagger$ Declination\\
		$^{*}$~http://irsa.ipac.caltech.edu,
		$^{**}$~\cite{2007A&A...474..653V}
	\end{center}
\end{table}

\section{ANALYSIS OF THE SYSTEM}\label{orbital_elements}
\subsection{Atmospheres modelling and the estimation of the physical parameters}\label{m1}
In spite of the fact that  the duplicity of the system was detected using high resolution spectroscopy and speckle interferometry, the system is seen as a single star even with the aid of the biggest telescopes. So, in order to estimate the physical parameters of the individual components of the system, we followed  Al-Wardat's complex method for analyzing CVBS.
The synthetic spectral energy distributions (SEDs) of the individual components of the system are computed using ATLAS9 line blanketed model atmospheres of ~\cite{1994KurCD..19.....K}  using a special subroutine. In order to build the model atmospheres of each components, we need preliminary input parameters ($T_{eff}$ and $\log g$), these are calculated as follows:

Using the apparent visual magnitude of the system $ m_v=6\fm60$ from the previous data (Table~\ref{tlab2}), and the visual magnitude difference $\triangle m=0\fm 33\pm0.06$ between the two components as the average of fifteen  $\triangle m$ measurements
of the filters $ \lambda$ $503$ -$ 850 $ (Table~\ref{deltam1}), we  calculated a preliminary individual  $m_v$ for each component using the following equations:
\begin{eqnarray}\label{Ma}
	m_A=m_v+2.5\log(1+10^{-0.4\Delta m}),\\
	m_B=m_A+\Delta m,
\end{eqnarray}
\noindent which give:

$m_v^A=7\fm20\pm0.03 , m_v^B=7\fm53\pm0.07$.

Combining these magnitudes  with the Hipparcos trigonometric parallax $(\pi_{Hip})$ from
\cite{2007A&A...474..653V}, we can derive the preliminary absolute magnitudes for the components using the following relation:
\begin{eqnarray}
	\label{eq3}
	\ M_V=m_v+5-5\log(d)-A_v
\end{eqnarray}
as follows: $M_V^A=5\fm85\pm0.04$ and $M_V^B=6\fm18\pm0.07$,
where $\ A_v$ is the interstellar reddening which was taken from NASA/IPAC (See Table~\ref{tlab2}).
\begin{table}[!ht]
	\begin{center}
		\caption{Magnitude difference between the components of the
			system Gliese 762.1, along with filters used to obtain the observations. }
		\label{deltam1}
		\begin{tabular}{lccc}
			\noalign{\smallskip}
			\hline\hline
			$\triangle m $& {$\sigma_{\Delta m}$}& filter ($\lambda/\Delta\lambda$)& references  \\
			\hline
			$0\fm26$ &   0.05  & $545nm/30 $&  1   \\						
			$0\fm33$ &  0.07   & $545nm/30 $&  2   \\		
			$0\fm27$ &   0.04  &$610nm/20 $&  3    \\
			$0\fm27$ &  0.15  & $648nm/41$ &   4  \\		
			$0\fm32$ &  0.15  &$503nm/40 $ &  4  \\		
			$0\fm24$ &  0.02  &$545nm/30 $&   5  \\			
			$0\fm25$ & 0.19  & $600nm/30 $  & 5  \\
			$0\fm24$ &  0.31  & $850nm/75 $  & 5  \\
			$0\fm41 $ &  -   & $698nm/39 $  &    6     \\
			$0\fm29$ &   0.03  & $600nm/30 $  & 7  \\
			$0\fm73$ &  -   & $698nm/39 $  &  6        \\
			$0\fm53$ &   -   & $550nm/40 $  &   6       \\
			$0\fm27$ &   -   & $754nm/44 $  &    6      \\
			$0\fm19$ &   -  & $562nm/40 $  &    8      \\
			$0\fm28$ &   -  & $692nm/40 $  &     8     \\
			\hline\hline
		\end{tabular}
		\\
		$^1${\cite{2005A&A...431..587P}},
		$^2${\cite{2002A&A...385...87B}},
		$^3${\cite{2004A&A...422..627B}},
		$^4${\cite{2004AJ....127.1727H}},
		$^5${\cite{2006BSAO...59...20B}},
		$^6${\cite{2008AJ....136..312H}},
		$^7${\cite{2007AstBu..62..339B}},
		$^8${\cite{2011AJ....141...45H}}.
	\end{center}
\end{table}
\begin{table}
	\begin{center}
		\centering
		\caption{ Relative position measurements obtained using different methods, which are used to build the orbit of the system.
			These points are taken from the Fourth Catalog of Interferometric Measurements of Binary Stars.}
		\small
		\label{po}
		\centering
		\begin{tabular}{lrcr}
			\hline\hline
			\multicolumn{1}{c}{Epoch} &
			\multicolumn{1}{c}{$\theta$}(deg) &
			\multicolumn{1}{c}{$\rho$}(arcsec) &
			References
			
			\\
			\hline
			\centering
			1980.4797 & 254.2  & 0.117  &    McA1983 (1)   \\
			1980.7228  & 226.7 & 0.106  &    McA1983 (1)	  \\
			1981.4736  & 310.4 & 0.081  &    McA1984a  (2)\\
			1983.4175 & 225.8 &  0.115  &    McA1987b  (3)   \\
			1984.7039 & 235.7&  0.112   &    McA1987b  (3) \\
			1985.4900 & 322.4  &  0.066  &   McA1987b (3)   \\
			1985.7390 &  273.1  & 0.104   &  Tok1988 (5)    \\
			1986.8883 & 308.5  &  0.065  &   McA1989(6)   \\
			1987.7618 & 172.7  &  0.067  &   McA1989(6)    \\
			1989.7059 & 286.0  & 0.093    &  Hrt1992b(7)  \\
			1989.8041 & 272.2  &  0.100   &  Bag1994(8)  \\
			1989.8096 & 268.2  &  0.102    & Bag1994(8)  \\
			1990.4322 & 171.6  & 0.062    &  Ism1992(9)    \\
			1991.25  & 284.0  &  0.100   &    HIP1997a(10)       \\
			1993.8438 & 271.3  & 0.102    &  Bag1994(8)     \\
			1994.7080 & 292.1 &   0.062   &  Hrt2000a(11)    \\
			1995.4397 & 244.8  &  0.113   &  Hrt1997(12)   \\
			1995.7621 & 201.9 &   0.086   &  Hrt1997(12) \\
			1996.6903 & 255.0 &   0.111   &  Hrt1997(12)  \\
			1999.8179 & 204.1 &  0.086    & Bag2002(13) \\
			2000.6166 & 271.9 &   0.099   &  Bag2004(14)\\
			2000.7640 & 253.8 &   0.111   &  Hor2002a(15)  \\
			2001.7550 & 134.6 &  0.065    &  Bag2006b(16)  \\
			2003.6339 & 235.6 &  0.114    &  Hor2008(17) \\
			2004.8150 & 75.0 &  0.110    &   Bag2007b(18)  \\
			2005.7662 & 153.5 &  0.054    & CIA2010(19)  \\
			2006.4381 & 44.9 &   0.104   &   Bag2013(20)  \\
			2006.5198 & 211.3 &   0.098   &  Hor2008(17)  \\
			2006.5227 & 213.8 &   0.095   &  Hor2008(17)  \\
			2007.8011 & 46.0 &   0.112   &   Hrt2009(21)  \\
			2010.4816 & 227.6 &   0.106   & Hor2011(22)  \\
			\hline\hline
		\end{tabular}
		\smallskip\\
		$^1${\cite{1983ApJS...51..309M}},
		$^2${\cite{1984ApJS...54..251M}},
		$^3${\cite{1987AJ.....93..688M}},
		$^4${\cite{1985A&AS...61..483T}},
		$^5${\cite{1988A&AS...72..563T}},
		$^6${\cite{1989AJ.....97..510M}},
		$^7${\cite{1992AJ....104..810H}},
		$^8${\cite{1994A&AS..105..503B}},
		$^9${\cite{1992A&AS...96..375I}},
		$^{10}${\cite{1997yCat.1239....0E}},
		$^{11}${\cite{2000AJ....119.3084H}},
		$^{12}${\cite{1997AJ....114.1639H}},
		$^{13}${\cite{2002A&A...385...87B}},
		$^{14}${\cite{2004A&A...422..627B}},
		$^{15}${\cite{2002AJ....123.3442H}},
		$^{16}${\cite{2006BSAO...59...20B}},
		$^{17}${\cite{2008AJ....136..312H}},
		$^{18}${\cite{2007AstBu..62..339B}},
		$^{19}${\cite{2010AJ....139.2308F}},
		$^{20}${\cite{2013AstBu..68...53B}},
		$^{21}${\cite{2009AJ....138..813H}},
		$^{22}${\cite{2011AJ....141...45H}}.
	\end{center}
\end{table}

The bolometric corrections$(B.C.)$, bolometric magnitudes and the stellar luminosities of the system were taken from \cite{1992adps.book.....L} and~\cite{2005oasp.book.....G}. These values, along with the following two equations:
\begin{eqnarray}
	\label{eq8}
	\log(R/R_\odot)= 0.5 \log(L/L_\odot)-2\log(T_{eff}/T_\odot),\\
	\label{eq5}
	\log g = \log(M/M_\odot)- 2\log(R/R_\odot) + 4.43
\end{eqnarray}

\noindent  were used to calculate the preliminary input parameters as:
$T_{eff}^A=5300K, T_{eff}^B=5050K$, $\log g_{A}=4.56, \log g_{B}=4.54$ and $R_{A}=0.815R_\odot$, $R_{B}=0.806R_\odot$.
$T_\odot$ were taken as $5777\rm{K}$.

The entire synthetic SED as if it is received from the system and measured above the earth's atmosphere is calculated using the following equations:
\begin{eqnarray}
	\label{eq6}
	F_\lambda \cdot d^2 = H_\lambda ^A \cdot R_{A} ^2 + H_\lambda ^B
	\cdot R_{B} ^2,
\end{eqnarray}
\noindent from which
\begin{eqnarray}
	\label{eq7}
	F_\lambda  = (R_{A} /d)^2(H_\lambda ^A + H_\lambda ^B \cdot(R_{B}/R_{A})^2) ,
\end{eqnarray}
\noindent
where $ R_{A}$ and $ R_{B}$ are the radii of the primary and secondary components of the system in solar units, $H_\lambda ^A $ and  $H_\lambda ^B$ are the fluxes at the surface of the star and $F_\lambda$ is the flux for the entire SED of the system above  the Earth's atmosphere which is located at a distance d (pc) from the system.

\begin{figure}[!ht]
	\centering
	\includegraphics[angle=0,width=14cm]{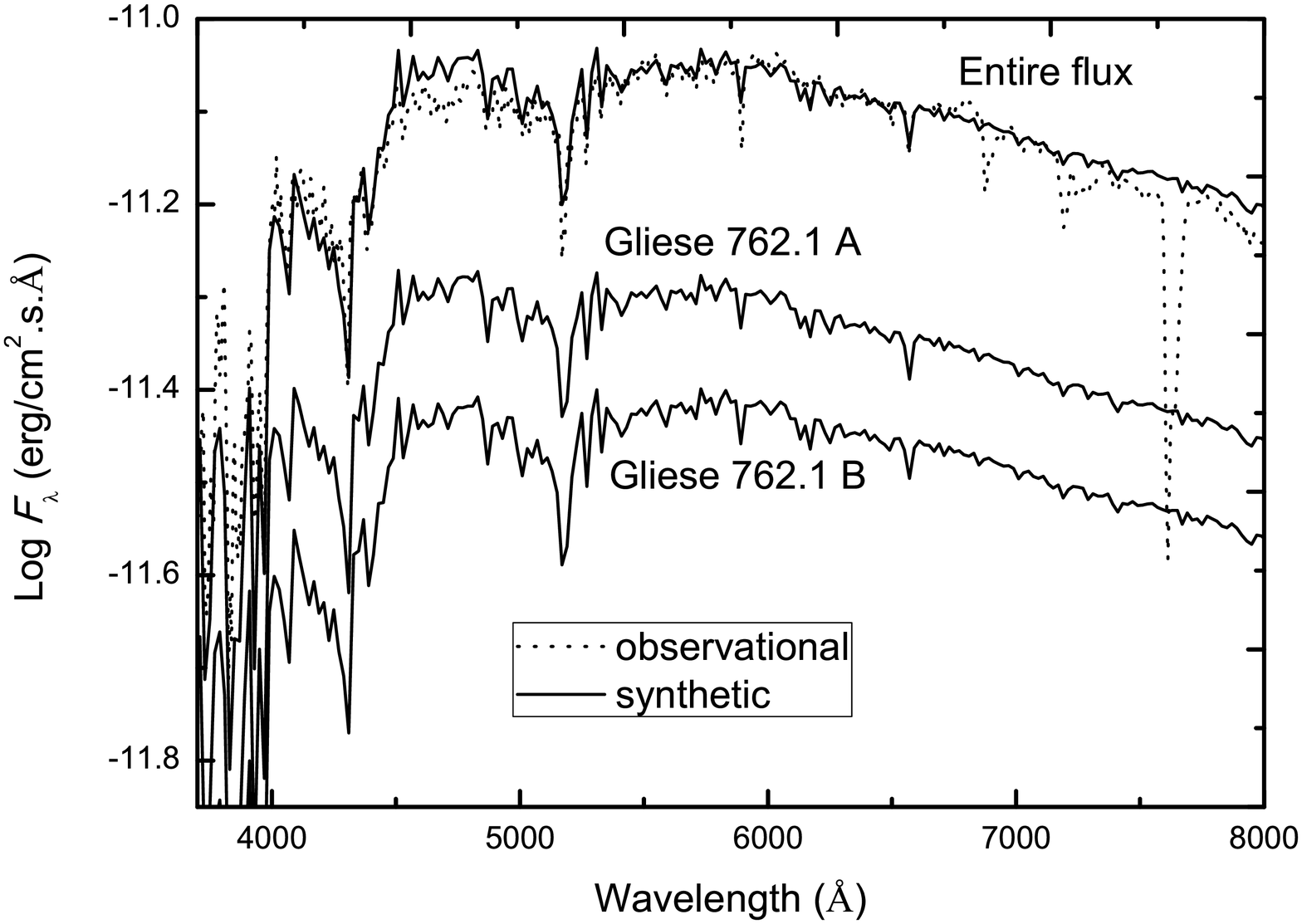}
	\caption{Best fit between the observed entire spectrum (dotted line) which was taken from~\citep{2002BSAO...54...29A} and the synthetic entire SED (solid line) for the system Gliese 762.1.  Individual synthetic SEDs were computed using  $T_{\rm eff}^A =5300\pm50$\,K, log $g_A=4.56\pm0.10, R_A=0.815\pm0.09R_\odot$, $T_{\rm eff}^B =5150\pm50$\,K, log $g_B=4.54\pm0.10$ and $R_B=0.806\pm0.10 R_\odot $, with $d=16.96$\ pc ($\pi=58.96\pm0.65 $ mas).} \label{fig1}
\end{figure}

The exact physical parameters of the components of the system are those which lead to the best fit between the entire synthetic SED and the observational one, which was taken form ~\cite{2002BSAO...54...29A}. The observational spectrum (Fig.~\ref{fig1}) was  obtained using  a low resolution grating ($325/4^{\circ}$ grooves/mm, {\AA}/px reciprocal dispersion)  within the UAGS spectrograph at the 1m (Zeiss-1000) SAO-Russian telescope.

Beside the visual best fit between the two spectra, the synthetic magnitudes, color indices and  line profiles especially those of Hydrogen $\ H_{\beta}$(4861.33\AA), $\  H_{\gamma}$(4340.5\AA) and $\ H_{\delta}$(4101\AA) should fit the observational ones. Otherwise, a new set of parameters would be tested in iterated way until the best fit is reached.

The best fit (Fig.~\ref{fig1}) was achieved using the  parameters shown in Table~\ref{tablef1}. The  luminosities and masses of the components were calculated using ~Equs.~\ref{eq8} and ~\ref{eq5}, and the spectral types of the components of the system were derived from \cite{1992adps.book.....L} empirical $Sp-M_V$ relation for main sequence stars.

\begin{table}[!ht]
	\small
	\begin{center}
		\caption{Fundamental parameters of the components of the system Gliese 762.1.}
		\label{tablef1}
		\begin{tabular}{lcc}
			\noalign{\smallskip}
			\hline\hline
			\noalign{\smallskip}
			Parameters & Comp. A &  Comp. B  \\
			\hline
			\noalign{\smallskip}
			$T_{\rm eff}$\,(K) & $5300\pm50$ & $5150\pm50$ \\
			Radius (R$_{\odot}$) & $0.845\pm0.09$ & $0.795\pm0.10$ \\
			$\log g$ & $4.52\pm0.10$ & $4.54\pm0.15$ \\
			$L (L_\odot)$ & $0.51\pm0.08 $  & $0.40\pm0.07$\\
			$M_{bol}$ & $5\fm48\pm0.90$ & $5\fm74\pm1.02$\\
			$M_{V}$ & $5\fm85\pm0.80$ & $6\fm18\pm0.85$\\
			Mass ($M_{\odot})^{*}$& $0.89 \pm0.08$ & $0.83 \pm0.07$  \\
			Sp. Type$^{**}$ & K0 & K1.5 \\
			\hline
			\multicolumn{1}{l}{Parallax (mas) }
			& \multicolumn{2}{c}{$58.96 \pm 0.65 $}\\

			\multicolumn{1}{l}{($\frac{M_A+M_B }{M_{\odot}}$)$^{***}$ }
			& \multicolumn{2}{c}{$1.72 \pm 0.60 $}\\

			\multicolumn{1}{l}{Age (Gy) }
		& \multicolumn{2}{c}{ $9\pm 1$}\\
			\hline\hline
			\noalign{\smallskip}
		\end{tabular}\\
		$^{*}${depending on the equation ~\ref{eq5}},\\
		$^{**}${depending on the tables of \cite{1992adps.book.....L}},\\
		$^{***}${depending on the orbital solution}.	
	\end{center}
\end{table}

\subsection{Orbital solution and Masses}\label{2.2}

Once available, the orbital elements of a binary system would enhance and help in examining the physical parameters of its individual components. The mass sum of the two components given by  equations ~\ref{eq31} and ~\ref{eq32} should coincide with that estimated from their positions on the evolutionary tracks and that calculated using the empirical equations and standard tables.

We followed Tokovinin's method \cite{1992ASPC...32..573T} to calculate the  orbital elements. The method  performs a least-squares adjustment to all available radial velocity and relative position observations, with weights inversely proportional to the square of their standard errors. The orbital solution involves: the orbital period, P; the semi-amplitudes of the primary and secondary velocities, K1 and K2; the eccentricity, e; the semi-major axis, a; the center of mass velocity, $\gamma$; and the time of primary minimum, $\ T_0$. The radial velocities for the system were taken from~\cite{1983PASP...95..201M}.

\begin{figure}[!ht]
	\centering
	\includegraphics[angle=0,width=14cm]{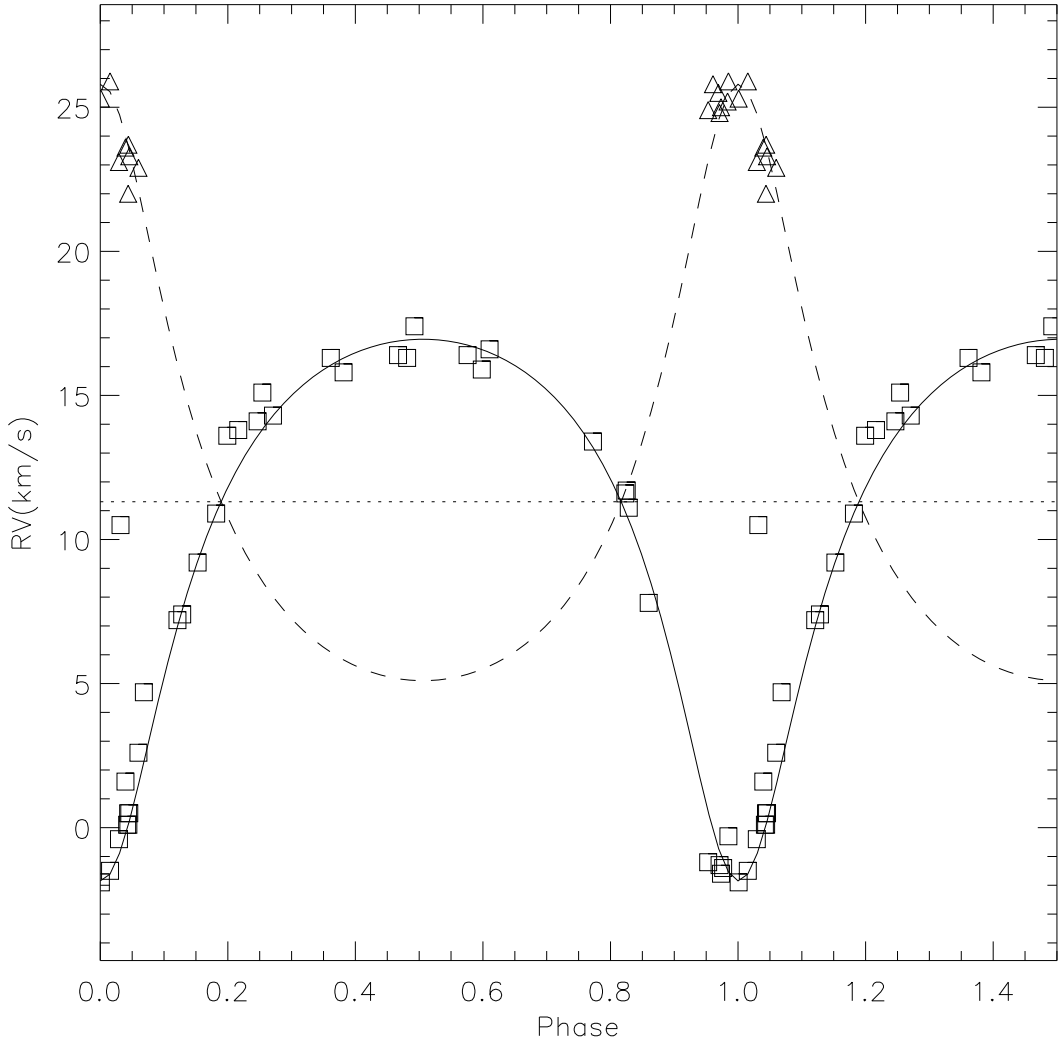}
	\caption{ Spectroscopic orbital solution for Gliese 762.1 in Table~\ref{orbit1} and radial velocities. Triangles represent radial velocities of the primary and squares represent radial velocities of the secondary component. The dotted line in the figure represents the center of mass velocity ( $ V_\gamma$=$ 11.31\pm0.12$ km$s^{-1}$). }
	\label{fig2}
\end{figure}

\begin{figure}[!ht]
	\centering
	\includegraphics[angle=0,width=14cm]{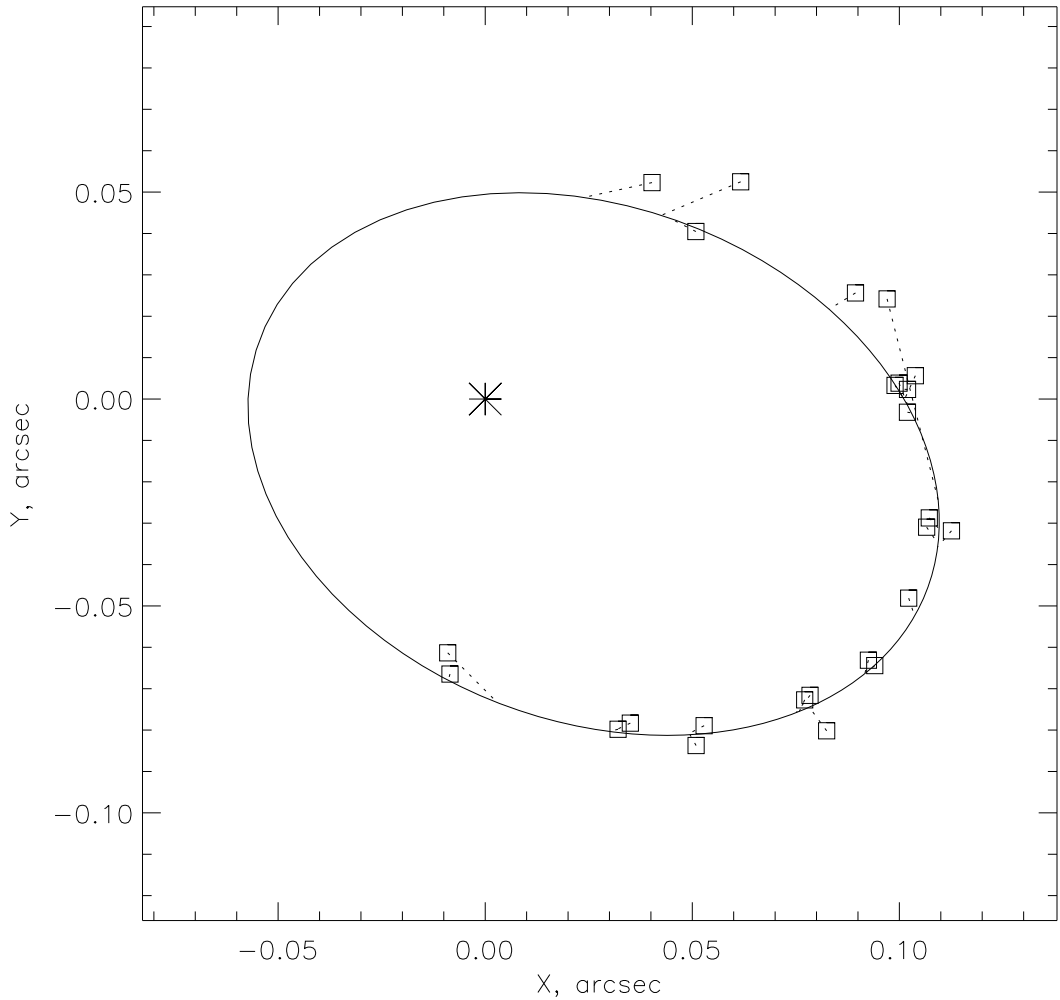}
	\caption{The best visual orbit of the system with the relative position measurements from the Fourth Catalog of Interferometric Measurements of Binary Stars. The squares represent the position of the primary component.}
	\label{fig3}
\end{figure}

Table~\ref{orbit1} lists the results of  the radial-velocity solution (Fig.~\ref{fig2}). The best orbit passes through the relative position measurements is shown in Fig.~\ref{fig3} and the resulting orbital elements are compared with earlier studies in Table~\ref{orbit}.

\begin{table}[!ht]
	\begin{center}
		\caption{Orbital solution of the system using the velocity curves Fig.~\ref{fig2}.}
		\label{orbit1}
		\begin{tabular}{lcc}
			\noalign{\smallskip}
			\hline\hline
			Parameters  & \cite{1983PASP...95..201M}  &  This paper
			\\
			\hline
			$P$, yr       &  $1.3477 \pm 0.0038$& $1.3534 \pm 0.0010$
			\\
			$T_0,$ MJD    &   44194.3 $\pm 3.2$& $47670.20 \pm 2.01$
			\\
			$e$          &   $0.416 \pm 0.016$ &    $0.40 \pm 0.013$
			\\
			$\omega$, deg   &   $177.1 \pm 2.70$ &  $179.0 \pm 2.24$
			\\
			$ K1$, km$s^{-1}$ & $ 9.52\pm0.16$&  $ 9.40\pm0.12$
			\\
			$ K2$, km$s^{-1}$ & $ 10.46\pm0.19$&  $ 10.35\pm0.19$
			\\
			$ V_\gamma$, km$s^{-1}$ & $ 11.41\pm0.014$&  $ 11.31\pm0.12$
			\\
			\hline\hline
		\end{tabular}\\
	\end{center}
\end{table}

\begin{table*}[!ht]
	\begin{center}
		\caption{Orbital elements, parallax and total mass of the system Gliese 762.1 using relative position measurements.}
		\label{orbit}
		\begin{tabular}{lcccc}
			\noalign{\smallskip}
			\hline\hline
			\noalign{\smallskip}
			Parameters   & \cite{2000IAUS..200P.135A} & \cite{2000AAS..145..215P}
			& \cite{2010AJ....139.2308F} & This work
			\\
			\hline
			$P$, yr      & $1.35458 \pm 0.00131$   & $1.352776 \pm 0.00071$   & $1.35297 \pm 0.00159$ & $1.3534 \pm 0.00075$
			\\
			$T_0,$ MJD   & $48641.21 \pm 3.10$     &  $46164.9 \pm 1.66$       & $46671.4 \pm 8.5$    &   $47670.20 \pm 2.37$
			\\
			$e$          & $0.340 \pm 0.013$       &  $0.3600 \pm 0.0078$      & $0.371 \pm 0.006$    &   $0.36 \pm 0.020$
			\\
			$a, $ arcsec & $0.084\pm 0.003 $       & $0.0860\pm 0.0014 $        & $0.0842 \pm 0.3 $   &  $0.0865 \pm 0.010 $
			\\
			$i $, deg    & $ 144.6 \pm 1.7$        & $ 144 \pm 2.4$             & $ 144.0 \pm 1.29$   &  $ 140.0 \pm 2.00$
			\\
			$\omega$, deg& $177.8 \pm 2.1$         & $356 \pm 2.1$              & $16.57 \pm 4.1$     &   $198.0 \pm 4.4$
			\\
			$\Omega$, deg& $74.6 \pm 6.8$          &   $243\pm 1.5$            & $256.9 \pm 2.666$    &  $253 \pm 6.55$
			\\
			$\pi$, mas   & $ 57.99\pm0.57$         & $ 57.3\pm0.3$             & $ 59.2\pm2.04$       & $ 58.96\pm0.65^{a}$
			\\
			$ M, M_\odot$& $1.62\pm0.18$           & $1.67\pm0.83$            & $1.59\pm0.18$         & $1.72\pm0.60$
			\\
			\hline\hline
		\end{tabular}\\
		${^a}$ New Hipparcos \cite{2007A&A...474..653V} (See Table~\ref{tlab2}).
	\end{center}
\end{table*}

The estimated orbital elements, semi-major axis, orbital period (see Table~\ref{orbit}), Hipparcos parallax of ~\cite{2007A&A...474..653V} as $\pi=58.96\pm0.65$ mas, along with Kepler's third law:
\begin{eqnarray}
\label{eq31}
\ M_A +M_B=(\frac{a^3}{\pi^3P^2})\ M_\odot
\end{eqnarray}
\begin{eqnarray}
\label{eq32}
\frac{\sigma_M }{M} =\sqrt{(3\frac{\sigma_\pi}{\pi})^2+(3\frac{\sigma_a}{a})^2+(2\frac{\sigma_p}{p})^2}
\end{eqnarray}
yield a mass sum with its corresponding error for the system as $\ M_A +M_B$=$1.72\pm0.60M_\odot$.
Using atmospheric modeling equation~\ref{eq5}, the total mass of the system is $1.69\pm0.22M_\odot$.

\section{SYNTHETIC PHOTOMETRY}
The entire and individual synthetic magnitudes are calculated by integrating the model fluxes over each bandpass of the system calibrated to the reference star (Vega) using the following equation \cite{2007ASPC..364..227M,2012PASA...29..523A}:
\begin{equation}
m_p[F_{\lambda,s}(\lambda)] = -2.5 \log \frac{\int P_{p}(\lambda)F_{\lambda,s}(\lambda)\lambda{\rm d}\lambda}{\int P_{p}(\lambda)F_{\lambda,r}(\lambda)\lambda{\rm d}\lambda}+ {\rm ZP}_p\,,
\end{equation}
 where $m_p$ is the synthetic magnitude of the passband $p$, $P_p(\lambda)$ is the dimensionless sensitivity function of the passband $p$, $F_{\lambda,s}(\lambda)$ is the synthetic SED of the object and $F_{\lambda,r}(\lambda)$ is the SED of Vega.  Zero points (ZP$_p$) from \cite{2007ASPC..364..227M} (and references there in) were adopted.

 The results of the calculated magnitudes and color  indices (Johnson: $U$, $B$, $ V$, $R$, $U-B$, $B-V$, $V-R$; Str\"{o}mgren: $u$, $v$, $b$,
 $y$, $u-v$, $v-b$, $b-y$ and Tycho: $B_{T}$, $ V_{T}$, $B_{T}-V_{T}$) of the entire
 system and individual components, in different photometrical systems,  are shown in Table~\ref{synth1}.

 \begin{table}
 	\small
 	\begin{center}
 		\caption{ Magnitudes and color indices  of the synthetic spectra of the  system Gliese 762.1.}
 		\label{synth1}
 		\begin{tabular}{lcccc}
 			\noalign{\smallskip}
 			\hline\hline
 			\noalign{\smallskip}
 			Sys. & Filter & entire & comp.& comp.\\
 			&     & $\sigma=\pm0.02$&   A    &     B      \\
 			\hline
 			\noalign{\smallskip}
 			Joh-  & $U$             & 7.99   & 8.53  & 9.01 \\
 			Cou.          & $B$     & 7.47   &  8.05 & 8.43  \\
 			& $V$                   & 6.60   &  7.20 &  7.53 \\
 			& $R$                       & 6.12   &  6.74 & 7.03  \\
 			&$U-B$                  & 0.52   & 0.48  & 0.58 \\
 			&$B-V$                  & 0.87   &  0.85 &  0.90 \\
 			&$V-R$                  & 0.47   &  0.46 & 0.50 \\
 			\noalign{\smallskip}
 			Str\"{o}m.        & $u$ & 9.15   & 9.69  &  10.18  \\
 			& $v$                   & 7.96   & 8.52  & 8.94  \\
 			& $b$                   & 7.06   & 7.65  &  8.00 \\
 			&  $y$                  & 6.56   & 7.16  & 7.48  \\
 			&$u-v$                  & 1.19   & 1.16  & 1.24 \\
 			&$v-b$                  & 0.90   & 0.87  & 0.94 \\
 			&$b-y$                  & 0.50   & 0.49  & 0.52 \\
 			\noalign{\smallskip}
 			Tycho       &$B_T$      & 7.71   & 8.28 & 8.68   \\
 			&$V_T$                  & 6.70   & 7.29 & 7.63  \\
 			&$B_T-V_T$              & 1.02   & 0.99 & 1.05\\
 			\hline\hline
 			\noalign{\smallskip}
 		\end{tabular}
 	\end{center}
 \end{table}

 \begin{table}
 	\small
 	\begin{center}
 		\caption{ Comparison between the observational and synthetic
 			magnitudes, colors and magnitude differences of the system
 			Gliese 762.1.} \label{synth3565}
 		\begin{tabular}{lcc}
 			\noalign{\smallskip}
 			\hline\hline
 			\noalign{\smallskip}
 			& Observed $^\dag$ & Synthetic (This work) \\
 			\hline
 			\noalign{\smallskip}
 			$V_{J}$ & $6\fm60$ & $6\fm60\pm0.02$\\
 			$B_{J}$ & $7\fm46$ & $7\fm47\pm0.02$\\
 			$R_{J}$ & $6\fm10$ & $6\fm12\pm0.02$\\
 			$B_T$  & $7\fm71\pm0.01$   &$7\fm71\pm0.02$\\
 			$V_T$  & $6\fm71\pm0.01$   &$6\fm70\pm0.02$\\
 			$(B-V)_{J}$&$ 0\fm86\pm0.01$ &$ 0\fm87\pm0.02$\\
 			$(U-B)_{J}$&$ 0\fm52\pm0.02$ &$ 0\fm52\pm0.02$\\
 			$\triangle m$  &$ 0\fm33^{\ddag}\pm0.06$  &$ 0\fm33\pm0.04$\\
 			\hline\hline \noalign{\smallskip}
 		\end{tabular}\\
 		$\dag$ See Table~\ref{tlab2}\\
 		$\ddag$ Average value for fifteen $\triangle m$ measurements (See Table~\ref{deltam1}).
 	\end{center}
 \end{table}

\begin{figure}
	\centering
	\includegraphics[angle=0,width=14cm]{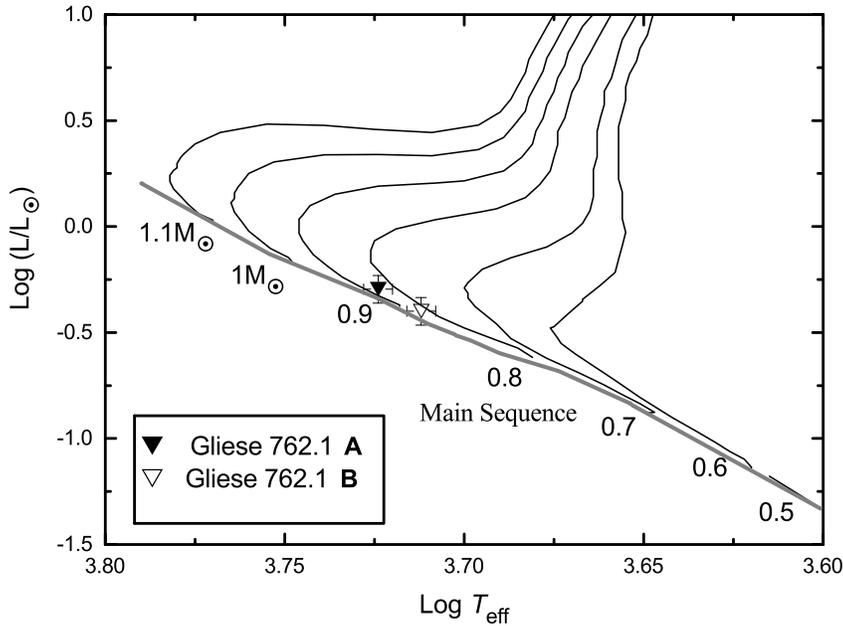}
	\caption{The  systems' components  on the evolutionary tracks of masses  ( 0.5, 0.6, 0.7,...., 1.1 $M_\odot$) of~\cite{2000yCat..41410371G}. }
	\label{fig4}
\end{figure}

\begin{figure}[!ht]
\resizebox{\hsize}{!} {\includegraphics[]{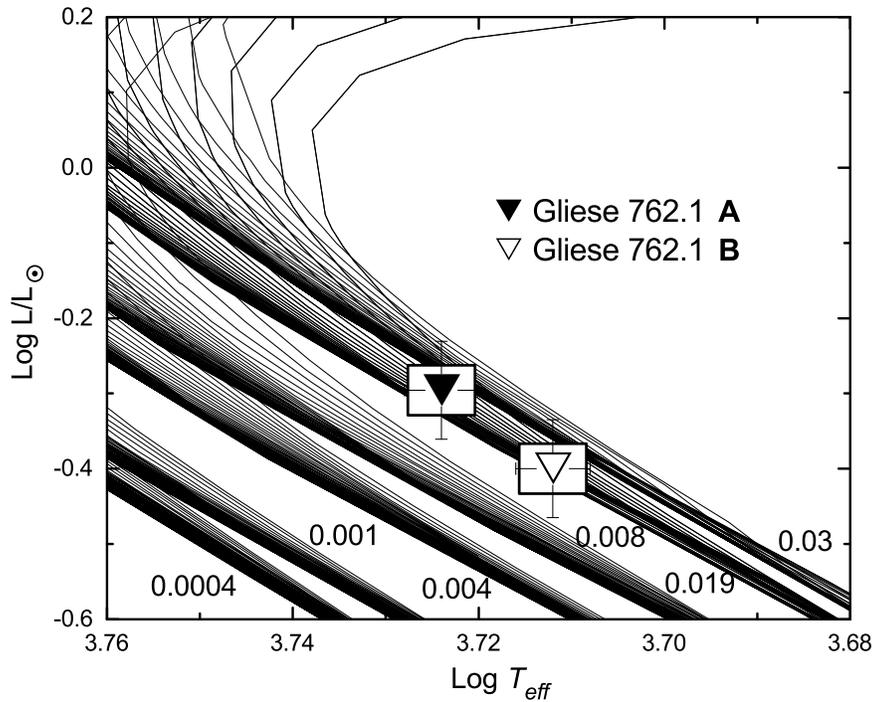}}
 \caption{The systems' components on the isochrones for low- and intermediate-mass stars of different metallicities of  ~\cite{2000A&AS..141..371G}.}
 \label{fig5}
\end{figure}

\begin{figure}[!ht]
\resizebox{\hsize}{!} {\includegraphics[]{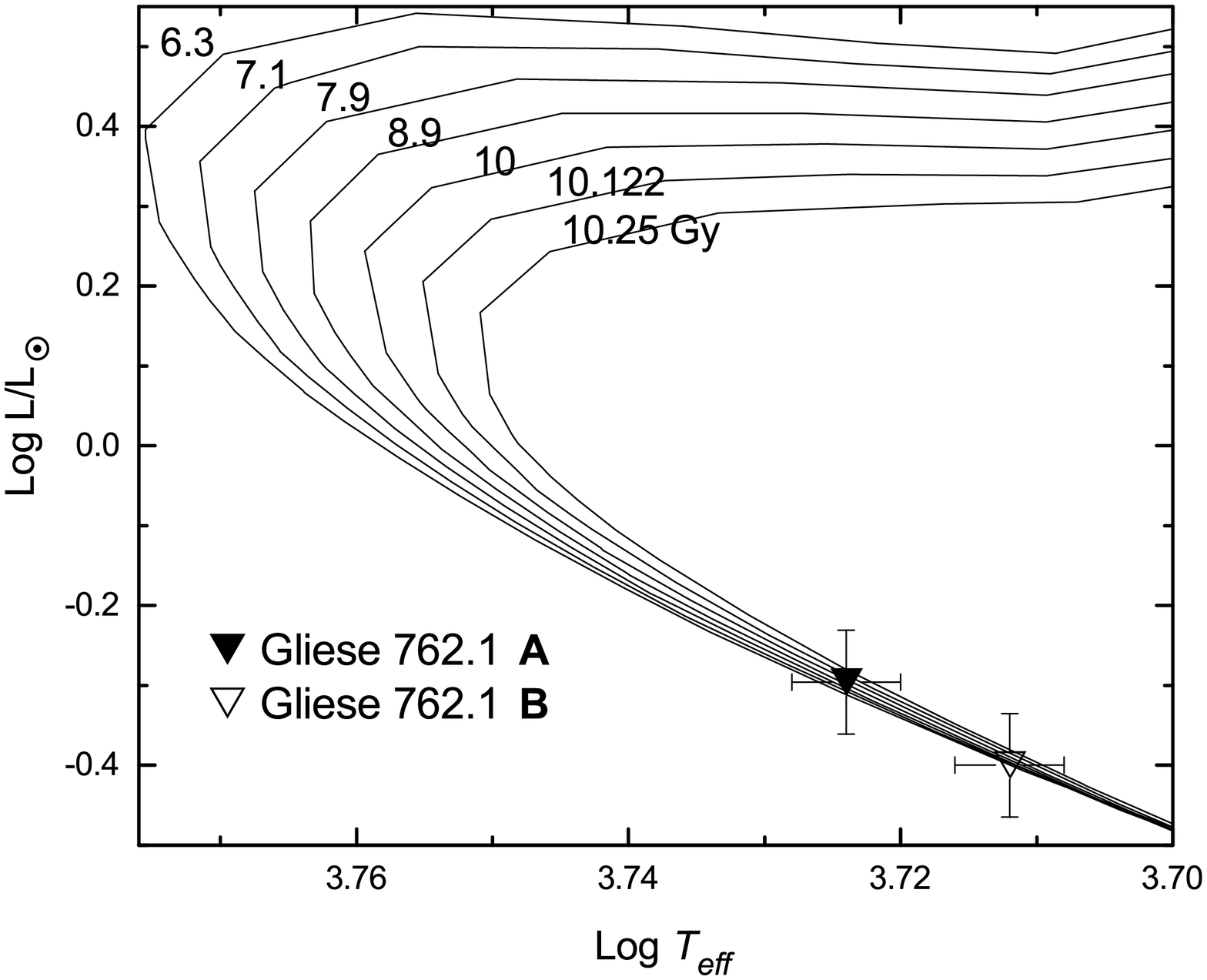}}
 \caption{The systems' components on the isochrones of low- and intermediate-mass, solar composition [$Z$=0.019, $Y$=0.273] stars of ~\cite{2000A&AS..141..371G}.}
 \label{fig6}
\end{figure}

\section{RESULTS AND DISCUSSION}

Table~\ref{synth3565} shows a high  consistency between  the  synthetic magnitudes and colors  and the observational ones. This gives a good indication about the reliability of the  estimated parameters listed in Table~\ref{tablef1}. Also, the resulted magnitude difference,  individual magnitudes and absolute magnitudes (Tables ~\ref{tablef1}\&~\ref{synth1}) are consistent with the calculated ones as preliminary input parameters.

The positions of the system's components on the evolutionary tracks of   \cite{2000A&AS..141..371G} (Fig.~\ref{fig4}) show that both components, of mass between $0.8$ and $0.9 M_\odot$ for each of them, belong to the main-sequence stars, but both show a slight displacement from the zero-age main-sequence upwards. And their positions on   ~\cite{2000A&AS..141..371G} isochrones for low- and intermediate-mass stars of different metallicities and that of the solar composition [$Z=0.019, Y=0.273$] are shown in Figs ~\ref{fig5} \& ~\ref{fig6}, which give an age of the system around $9\pm 1$ Gy.

The spectral types and luminosity classes of both components are assigned as K0V and K1.5V for the primary and secondary components respectively, and their positions on the evolutionary tracks are showed in Fig.~\ref{fig4}, which are brighter than those given by ~\cite{2010AJ....139.2308F}  as K2V and K4V.

The estimated orbital elements of the system (Tables~\ref{orbit1}~\&~\ref{orbit}) are consistent with previous works. The orbit of the system was solved using a combination of the relative position measurements and the radial velocity curves, which gives more reliable and accurate results.

The mass sum of the system components and the individual masses were calculated and estimated in three different ways; using the physical parameters and standard relations as $M_a=0.89\pm0.08M_\odot, M_b=0.83\pm0.07M_\odot$,  using the  orbital elements with Hipparcos parallax as $M_a+M_b=1.72\pm0.60M_\odot$ and depending on the positions of the system's components on the evolutionary tracks of   \cite{2000A&AS..141..371G} (Fig.~\ref{fig4}) which coincide with the calculated ones.

Depending on the estimated parameters of the system's components and their positions on the evolutionary tracks,  fragmentation
is a possible  process for the formation of the system. Where \cite{1994MNRAS.269..837B} concludes that fragmentation of a rotating disk
around an incipient central protostar is possible, as long as
there is continuing infall.  \cite{2001IAUS..200.....Z} pointed out that
hierarchical  fragmentation during rotational collapse has been
invoked to produce binaries and multiple systems.

It is worthwhile to mention here that the system Gliese 762.1 is a detached one with an orbital period of $1.3534\pm0.0010$yr. So, this system is a visually close but not a contact binary, and if we compare it with  other extremely K-type close binary systems like BI Vulpeculae ~\citep{2013ApJS..209...13Q}, AD Cancri ~\citep{2007ApJ...671..811Q} and PY Virginis ~\citep{2013AJ....145...39Z},  which have shorter orbital periods (~days) and lower angular
momentum among K-type binary stars,  we find that the orbital evolution of such systems are affected by the existence of a third component by removing angular momentum from the central binary system during the early stellar formation process or/and later dynamical interactions. However, as for Gliese 762.1, such dynamical interactions may not exist. It may form directly from stellar formation process because the orbital separation between the two components is much larger.

\section{CONCLUSIONS}

We present the results of the complex analysis of the double-lined spectroscopic binary system Gliese 762.1.  We were able to achieve the best fit between the entire synthetic SED's and the  observational one (Fig.~\ref{fig1}) by producing and calibrating synthetic SED  of the individual components in an iterated method. The orbit of the system and its radial velocities were also solved to estimate reliable orbital elements consistent with the physical ones of atmospheres modeling.

We relayed on Hipparcos parallax ($58.96 \pm 0.65 $ mas, $d=16.96$\,pc \cite{2007A&A...474..653V}) for the calculations of the entire SED and the masses of the system. The Hipparcos parallax was not that far from the dynamical parallaxes introduced in previous orbital solutions (see Table~\ref{orbit}).

\normalem
\begin{acknowledgements}
This research has made use of SAO/NASA, SIMBAD database, Fourth Catalog of Interferometric Measurements of Binary Stars, IPAC
data systems and CHORIZOS code of photometric
and spectrophotometric data analysis.
We would like to thank J\"{u}rgen Weiprecht from Astrophysikalisches Institut und Universit\"{a}ts-Sternwarte, FSU Jena for his help. S. Masda would like to thank ministry of higher education and scientific research in Yemen for the scholarship as well as S. Masda would like to thank Hadhramout University (Faculty of Al-Mahra Education) in Yemen for facilitate some things.

\end{acknowledgements}


\begin{thebibliography}{56}
\providecommand\natexlab[1]{#1}
\providecommand\JournalTitle[1]{#1}

\bibitem[{Al-Wardat}(2002{\natexlab{a}})]{2002BSAO...53...51A}
{Al-Wardat}, M.~A. 2002{\natexlab{a}}, Bull.~Special Astrophys.~Obs., 53, 51

\bibitem[{Al-Wardat}(2002{\natexlab{b}})]{2002BSAO...54...29A}
{Al-Wardat}, M.~A. 2002{\natexlab{b}}, Bulletin of the Special Astrophysics
  Observatory, 54, 29

\bibitem[{Al-Wardat}(2007)]{2007AN....328...63A}
{Al-Wardat}, M.~A. 2007, Astronomische Nachrichten, 328, 63

\bibitem[{Al-Wardat}(2009)]{2009AN....330..385A}
{Al-Wardat}, M.~A. 2009, Astronomische Nachrichten, 330, 385

\bibitem[{Al-Wardat}(2012)]{2012PASA...29..523A}
{Al-Wardat}, M.~A. 2012, \pasa, 29, 523

\bibitem[{Al-Wardat}(2014)]{2014AstBu..69..454A}
{Al-Wardat}, M.~A. 2014, Astrophysical Bulletin, 69, 454

\bibitem[{Al-Wardat} {et~al.}(2014{\natexlab{a}})]{2014AstBu..69...58A}
{Al-Wardat}, M.~A., {Balega}, Y.~Y., {Leushin}, V.~V., {et~al.}
  2014{\natexlab{a}}, Astrophysical Bulletin, 69, 58

\bibitem[{Al-Wardat} {et~al.}(2014{\natexlab{b}})]{2014AstBu..69..198A}
{Al-Wardat}, M.~A., {Balega}, Y.~Y., {Leushin}, V.~V., {et~al.}
  2014{\natexlab{b}}, Astrophysical Bulletin, 69, 198

\bibitem[{Al-Wardat} \& {Widyan}(2009)]{2009AstBu..64..365A}
{Al-Wardat}, M.~A., \& {Widyan}, H. 2009, Astrophysical Bulletin, 64, 365

\bibitem[{Al-Wardat} {et~al.}(2014{\natexlab{c}})]{2014PASA...31....5A}
{Al-Wardat}, M.~A., {Widyan}, H.~S., \& {Al-thyabat}, A. 2014{\natexlab{c}},
  \pasa, 31, 5

\bibitem[{Arenou} {et~al.}(2000)]{2000IAUS..200P.135A}
{Arenou}, F., {Halbwachs}, J.-L., {Mayor}, M., {Palasi}, J., \& {Udry}, S.
  2000, in IAU Symposium, Vol. 200, IAU Symposium, 135P

\bibitem[{Arenou} {et~al.}(2002)]{2002EAS.....2..155A}
{Arenou}, F., {Halbwachs}, J.-L., {Mayor}, M., \& {Udry}, S. 2002, in EAS
  Publications Series, Vol.~2, EAS Publications Series, ed. O.~{Bienayme} \&
  C.~{Turon}, 155

\bibitem[{Balega} {et~al.}(2004)]{2004A&A...422..627B}
{Balega}, I., {Balega}, Y.~Y., {Maksimov}, A.~F., {et~al.} 2004, \aap, 422, 627

\bibitem[{Balega} {et~al.}(2006)]{2006BSAO...59...20B}
{Balega}, I.~I., {Balega}, A.~F., {Maksimov}, E.~V., {et~al.} 2006,
  Bull.~Special Astrophys.~Obs., 59, 20

\bibitem[{Balega} {et~al.}(1994)]{1994A&AS..105..503B}
{Balega}, I.~I., {Balega}, Y.~Y., {Belkin}, I.~N., {et~al.} 1994, \aaps, 105,
  503

\bibitem[{Balega} {et~al.}(2013)]{2013AstBu..68...53B}
{Balega}, I.~I., {Balega}, Y.~Y., {Gasanova}, L.~T., {et~al.} 2013,
  Astrophysical Bulletin, 68, 53

\bibitem[{Balega} {et~al.}(2002)]{2002A&A...385...87B}
{Balega}, I.~I., {Balega}, Y.~Y., {Hofmann}, K.-H., {et~al.} 2002, \aap, 385,
  87

\bibitem[{Balega} {et~al.}(2007{\natexlab{a}})]{2007AstBu..62..339B}
{Balega}, I.~I., {Balega}, Y.~Y., {Maksimov}, A.~F., {et~al.}
  2007{\natexlab{a}}, Astrophysical Bulletin, 62, 339

\bibitem[{Balega} {et~al.}(2007{\natexlab{b}})]{2007A&A...464..635B}
{Balega}, Y.~Y., {Beuzit}, J.-L., {Delfosse}, X., {et~al.} 2007{\natexlab{b}},
  \aap, 464, 635

\bibitem[{Batten} {et~al.}(1989)]{1989PDAO...17....1B}
{Batten}, A.~H., {Fletcher}, J.~M., \& {MacCarthy}, D.~G. 1989, Publications of
  the Dominion Astrophysical Observatory Victoria, 17, 1

\bibitem[{Bonnell}(1994)]{1994MNRAS.269..837B}
{Bonnell}, I.~A. 1994, \mnras, 269, 837

\bibitem[{Duquennoy} {et~al.}(1991)]{1991A&AS...88..281D}
{Duquennoy}, A., {Mayor}, M., \& {Halbwachs}, J.-L. 1991, \aaps, 88, 281

\bibitem[{ESA}(1997)]{1997yCat.1239....0E}
{ESA}. 1997, {The Hipparcos and Tycho Catalogues (ESA)}

\bibitem[{Farrington} {et~al.}(2010)]{2010AJ....139.2308F}
{Farrington}, C.~D., {ten Brummelaar}, T.~A., {Mason}, B.~D., {et~al.} 2010,
  \aj, 139, 2308

\bibitem[{Forveille} {et~al.}(1999)]{1999A&A...351..619F}
{Forveille}, T., {Beuzit}, J.-L., {Delfosse}, X., {et~al.} 1999, \aap, 351, 619

\bibitem[{Girardi} {et~al.}(2000{\natexlab{a}})]{2000A&AS..141..371G}
{Girardi}, L., {Bressan}, A., {Bertelli}, G., \& {Chiosi}, C.
  2000{\natexlab{a}}, \aaps, 141, 371

\bibitem[{Girardi} {et~al.}(2000{\natexlab{b}})]{2000yCat..41410371G}
{Girardi}, L., {Bressan}, A., {Bertelli}, G., \& {Chiosi}, C.
  2000{\natexlab{b}}, VizieR Online Data Catalog, 414, 10371

\bibitem[{Gray}(2005)]{2005oasp.book.....G}
{Gray}, D.~F. 2005, {The Observation and Analysis of Stellar Photospheres}, 505

\bibitem[{Hartkopf} \& {Mason}(2009)]{2009AJ....138..813H}
{Hartkopf}, W.~I., \& {Mason}, B.~D. 2009, \aj, 138, 813

\bibitem[{Hartkopf} {et~al.}(1992)]{1992AJ....104..810H}
{Hartkopf}, W.~I., {McAlister}, H.~A., \& {Franz}, O.~G. 1992, \aj, 104, 810

\bibitem[{Hartkopf} {et~al.}(1997)]{1997AJ....114.1639H}
{Hartkopf}, W.~I., {McAlister}, H.~A., {Mason}, B.~D., {et~al.} 1997, \aj, 114,
  1639

\bibitem[{Hartkopf} {et~al.}(2000)]{2000AJ....119.3084H}
{Hartkopf}, W.~I., {Mason}, B.~D., {McAlister}, H.~A., {et~al.} 2000, \aj, 119,
  3084

\bibitem[{Horch} {et~al.}(2011)]{2011AJ....141...45H}
{Horch}, E.~P., {Gomez}, S.~C., {Sherry}, W.~H., {et~al.} 2011, \aj, 141, 45

\bibitem[{Horch} {et~al.}(2004)]{2004AJ....127.1727H}
{Horch}, E.~P., {Meyer}, R.~D., \& {van Altena}, W.~F. 2004, \aj, 127, 1727

\bibitem[{Horch} {et~al.}(2002)]{2002AJ....123.3442H}
{Horch}, E.~P., {Robinson}, S.~E., {Meyer}, R.~D., {et~al.} 2002, \aj, 123,
  3442

\bibitem[{Horch} {et~al.}(2008)]{2008AJ....136..312H}
{Horch}, E.~P., {van Altena}, W.~F., {Cyr}, Jr., W.~M., {et~al.} 2008, \aj,
  136, 312

\bibitem[{Ismailov}(1992)]{1992A&AS...96..375I}
{Ismailov}, R.~M. 1992, \aaps, 96, 375

\bibitem[{Kurucz}(1994)]{1994KurCD..19.....K}
{Kurucz}, R. 1994, Solar abundance model atmospheres for 0,1,2,4,8 km/s.~Kurucz
  CD-ROM No.~19.~ Cambridge, Mass.: Smithsonian Astrophysical Observatory,
  1994., 19

\bibitem[{Lang}(1992)]{1992adps.book.....L}
{Lang}, K.~R. 1992, {Astrophysical Data I. Planets and Stars.}, 133

\bibitem[{Ma{\'{\i}}z Apell{\'a}niz}(2007)]{2007ASPC..364..227M}
{Ma{\'{\i}}z Apell{\'a}niz}, J. 2007, in Astronomical Society of the Pacific
  Conference Series, Vol. 364, The Future of Photometric, Spectrophotometric
  and Polarimetric Standardization, ed. C.~{Sterken} (San Francisco:
  Astronomical Society of the Pacific), 227

\bibitem[{McAlister} {et~al.}(1984)]{1984ApJS...54..251M}
{McAlister}, H.~A., {Hartkopf}, W.~I., {Gaston}, B.~J., {Hendry}, E.~M., \&
  {Fekel}, F.~C. 1984, \apjs, 54, 251

\bibitem[{McAlister} {et~al.}(1983)]{1983ApJS...51..309M}
{McAlister}, H.~A., {Hartkopf}, W.~I., {Hendry}, E.~M., {Campbell}, B.~G., \&
  {Fekel}, F.~C. 1983, \apjs, 51, 309

\bibitem[{McAlister} {et~al.}(1987)]{1987AJ.....93..688M}
{McAlister}, H.~A., {Hartkopf}, W.~I., {Hutter}, D.~J., \& {Franz}, O.~G. 1987,
  \aj, 93, 688

\bibitem[{McAlister} {et~al.}(1989)]{1989AJ.....97..510M}
{McAlister}, H.~A., {Hartkopf}, W.~I., {Sowell}, J.~R., {Dombrowski}, E.~G., \&
  {Franz}, O.~G. 1989, \aj, 97, 510

\bibitem[{McClure}(1983)]{1983PASP...95..201M}
{McClure}, R.~D. 1983, \pasp, 95, 201

\bibitem[{Pluzhnik}(2005)]{2005A&A...431..587P}
{Pluzhnik}, E.~A. 2005, \aap, 431, 587

\bibitem[{Pourbaix}(2000)]{2000AAS..145..215P}
{Pourbaix}, D. 2000, \aaps, 145, 215

\bibitem[{Pourbaix} \& {Jorissen}(2000)]{2000A&AS..145..161P}
{Pourbaix}, D., \& {Jorissen}, A. 2000, \aaps, 145, 161

\bibitem[{Qian} {et~al.}(2007)]{2007ApJ...671..811Q}
{Qian}, S.-B., {Yuan}, J.-Z., {Soonthornthum}, B., {et~al.} 2007, \apj, 671,
  811

\bibitem[{Qian} {et~al.}(2013)]{2013ApJS..209...13Q}
{Qian}, S.-B., {Liu}, N.-P., {Li}, K., {et~al.} 2013, \apjs, 209, 13

\bibitem[{Tokovinin}(1992)]{1992ASPC...32..573T}
{Tokovinin}, A. 1992, in Astronomical Society of the Pacific Conference Series,
  Vol.~32, IAU Colloq. 135: Complementary Approaches to Double and Multiple
  Star Research, ed. H.~A. {McAlister} \& W.~I. {Hartkopf}, 573

\bibitem[{Tokovinin}(1985)]{1985A&AS...61..483T}
{Tokovinin}, A.~A. 1985, \aaps, 61, 483

\bibitem[{Tokovinin} \& {Ismailov}(1988)]{1988A&AS...72..563T}
{Tokovinin}, A.~A., \& {Ismailov}, R.~M. 1988, \aaps, 72, 563

\bibitem[{van Leeuwen}(2007)]{2007A&A...474..653V}
{van Leeuwen}, F. 2007, \aap, 474, 653

\bibitem[{Zhu} {et~al.}(2013)]{2013AJ....145...39Z}
{Zhu}, L.~Y., {Qian}, S.~B., {Liu}, N.~P., {Liu}, L., \& {Jiang}, L.~Q. 2013,
  \aj, 145, 39

\bibitem[{Zinnecker} \& {Mathieu}(2001)]{2001IAUS..200.....Z}
{Zinnecker}, H., \& {Mathieu}, R., eds. 2001, IAU Symposium, Vol. 200, {The
  Formation of Binary Stars}

\end{thebibliography}

\end{document}